\documentclass[journal=jacsat]{achemso}
\usepackage{chemformula} % Formula subscripts using \ch{}
\usepackage[T1]{fontenc} % Use modern font encodings
\usepackage{amsmath,amssymb,amsfonts,latexsym,graphicx}
\usepackage{geometry}
\usepackage{indentfirst}
\usepackage{fancyhdr}
\usepackage{subfigure}
\usepackage{titletoc}
\usepackage{titlesec}
\usepackage{color,ulem}
\usepackage[version=3]{mhchem} % Formula subscripts using \ce{}
\setkeys{acs}{maxauthors = 0} % references of many authors
\usepackage{braket}
\usepackage{bm}
\usepackage{diagbox}
\usepackage{multirow}

\newcommand{\kb}[2]{\vert #1 \rangle \langle #2 \vert}

\author{Daeheum Cho}\email{daeheumc@uci.edu}
\affiliation{Department of Chemistry and Physics and Astronomy, University of California,
Irvine, California 92697-2025, USA}
\author{Jeremy R. Rouxel}
\affiliation{Department of Chemistry and Physics and Astronomy, University of California,
Irvine, California 92697-2025, USA}
\author{Shaul Mukamel}\email{smukamel@uci.edu}
\affiliation{Department of Chemistry and Physics and Astronomy, University of California,
Irvine, California 92697-2025, USA}
\author{Garnet Kin-Lic Chan}\email{gkc1000@gmail.com}
\affiliation{Division of Chemistry and Chemical Engineering, California Institute
of Technology, Pasadena, CA 91125, USA}
\author{Zhendong Li}\email{zhendongli2008@gmail.com}
\affiliation{Division of Chemistry and Chemical Engineering, California Institute of Technology, Pasadena, CA 91125, USA,\\
Key Laboratory of Theoretical and Computational Photochemistry, Ministry of Education, College of Chemistry, Beijing Normal University, Beijing 100875, China}

\title[An \textsf{achemso} demo]
  {Stimulated X-ray Raman and Absorption Spectroscopy of Iron-Sulfur
  Dimers}

\begin{document}
%%%%%%%%%%%%%%%%%%%%%%%%%%%%%%%%%%%%%%%%%%%%%%%%%%%%%%%%%%%%%%%%%%%%%
%% The manuscript does not need to include \maketitle, which is
%% executed automatically.  The document should begin with an
%% abstract, if appropriate.  If one is given and should not be, the
%% contents will be gobbled.
%%%%%%%%%%%%%%%%%%%%%%%%%%%%%%%%%%%%%%%%%%%%%%%%%%%%%%%%%%%%%%%%%%%%%

\begin{abstract}
Iron-sulfur complexes play an important role in biological processes such as metabolic electron transport.
A detailed understanding of   the mechanism of long range electron transfer requires knowledge of the electronic structure of the complexes, which has traditionally been challenging to obtain, either by theory or by experiment, but the situation has begun to change with advances in quantum chemical methods and intense free electron laser light sources.
We compute the signals from stimulated X-ray Raman spectroscopy (SXRS) and absorption spectroscopy of homovalent and mixed-valence [2Fe-2S] complexes, using the {\it ab initio} density matrix renormalization group (DMRG) algorithm.
The simulated spectra show clear signatures of the theoretically predicted dense low-lying excited states within the d-d manifold.
Furthermore, the difference in  signal intensity between the absorption-active and Raman-active states provides a potential mechanism to selectively excite states by a proper tuning of the excitation pump, to access the electronic dynamics within this manifold.
\end{abstract}

%%%%%%%%%%%%%%%%%%%%%%%%%%%%%%%%%%%%%%%%%%%%%%%%%%%%%%%%%%%%%%%%%%%%%
%% Start the main part of the manuscript here.
%%%%%%%%%%%%%%%%%%%%%%%%%%%%%%%%%%%%%%%%%%%%%%%%%%%%%%%%%%%%%%%%%%%%%

Iron-sulfur (Fe-S) complexes are pervasive in Nature\cite{beinert1997iron,howard1996structural,rees2003interface}. The most common motifs include the [2Fe-2S] and [4Fe-4S] clusters contained in ferredoxin (Fd) proteins, which mediate
electron transfer in many metabolic reactions.
During  electron transfer, the [2Fe-2S] clusters  convert between the homovalent
[2Fe(III,III)-2S] and mixed-valence
[2Fe(III,II)-2S] forms upon receiving or donating electrons. However,
unlike in many other biological systems, the detailed mechanism
of electron transfer involving Fe-S clusters is not well understood at the molecular level. This is often attributed to the rather complicated electronic structure of these clusters\cite{blumberger2015recent}.
Spectroscopic and magnetic susceptibility studies\cite{palmer1966magnetic,brintzinger1966ligand,dunham1971two,dunham1971structure} established early on
that in these complexes, each iron atom can be formally
viewed as a high-spin ferric iron with $S=5/2$ or ferrous iron with $S=2$, coordinated to four sulfur atoms (either from the thiolate or bridge sulfide) in a (distorted) tetrahedral environment. The ground states of these Fe-S clusters are formed by antiferromagnetically coupled Fe(III)-Fe(III)
and Fe(III)-Fe(II) pairs, respectively, which leads to
a diamagnetic $S=0$ state for the homovalent dimer and
a $S=1/2$ state for the mixed-valence dimer which has a clear
electron paramagnetic resonance (EPR) signature\cite{palmer1966magnetic,brintzinger1966ligand}.
Similar electronic features have been widely
observed in a variety of synthetic analogs of [2Fe-2S] clusters\cite{mayerle1973synthetic,mayerle1975synthetic,venkateswara2004synthetic}. While these
basic features can be described
by the Heisenberg double exchange model\cite{girerd1983electron,noodleman1984electronic} in combination with
broken-symmetry density functional theory (BS-DFT)\cite{noodleman1981valence,noodleman1985models,yamaguchi1989antiferromagnetic,noodleman2002insights},
more recent theoretical work\cite{sharma_low-energy_2014,chilkuri2019ligand} has shown that the excited state spectrum
is much more involved and cannot be described by this simple approach. Instead, using the {\it ab initio} density matrix renormalization
group (DMRG) algorithm\cite{white_ab_1999,chan_highly_2002,legeza2003controlling,moritz2005convergence,sharma_spin-adapted_2012},
it has been shown that the low-energy spectrum is very dense due to the presence of a large number of d-d excited states arising
from both orbital transitions and spin recouplings\cite{sharma_low-energy_2014}.
Indirect experimental evidence of the dense low-energy manifold has recently been obtained using iron L-edge 2p3d resonant inelastic X-ray scattering (RIXS)\cite{van2018electronic}.

In general, experimental access to the low-lying electronic excited states of [2Fe-2S] dimers through optical absorption is difficult,
because d-d ligand-field excitations are essentially electric-dipole forbidden\cite{eaton1971tetrahedral} as in mononuclear Fe-S complexes. In this work, we explore the use of nonlinear optical spectroscopies to probe the low-lying spectra of [2Fe-2S] dimers. Specifically, we compute the stimulated X-ray Raman spectroscopy (SXRS) signals of the homovalent and mixed-valent [2Fe-2S] complexes, and compare them to simulated absorption signals, using electronic excited states computed with the {\it ab initio} DMRG technique. We find that the absorption and SXRS techniques complement each other by accessing different parts of the electronic spectrum, and together can effectively probe the dense d-d electronic states in the Fe-S clusters. Thanks to the availability of accurate many-electron wavefunctions from DMRG, a detailed assignment of the signals then becomes possible,
to aid the understanding of experimental spectroscopy of iron-sulfur dimers in the future.

{\it Simulation of stimulated X-ray Raman (SXRS) signals} We shall calculate the stimulated X-ray Raman spectroscopy signals using the minimal coupling Hamiltonian rather than the multipolar Hamiltonian.
The minimal coupling field-matter interaction
Hamiltonian is given by~\cite{Chernyak15MinimalCoupling}
\begin{equation}
\hat{H}_{\text{int}}(t)=-\int d\boldsymbol{r} \hat{\boldsymbol{j}}(\boldsymbol{r})\cdot \hat{\boldsymbol{A}}(\boldsymbol{r})+\frac{1}{2}\int d\boldsymbol{r}\ \hat{\sigma}(\boldsymbol{r})\hat{\boldsymbol{A}}^{2}(\boldsymbol{r})\label{hint}
\end{equation}
where we work in atomic units and $\hat{\boldsymbol{j}}(\boldsymbol{r})$ and $\hat{\sigma}(\boldsymbol{r})$ are the current and charge density operators, respectively, and $\boldsymbol{A}(\boldsymbol{r})$ is the vector potential.
The current $\hat{\boldsymbol{j}}(\boldsymbol{r})$ and charge density $\hat{\sigma}(\boldsymbol{r})$ operators are defined as
%{\color{red}Are there any general references for these derivations?}
\begin{equation}
\hat{\boldsymbol{j}}(\boldsymbol{r})=\frac{1}{2i} \bigg( \hat{\psi}^{\dagger}(\boldsymbol{r}) \nabla \hat{\psi}(\boldsymbol{r}) -(\nabla \hat{\psi}^{\dagger}(\boldsymbol{r})) \hat{\psi}(\boldsymbol{r})\bigg)
\label{eq:current}
\end{equation}
\begin{equation}
\hat{\sigma}(\boldsymbol{r})=\hat{\psi}^{\dagger}(\boldsymbol{r}) \hat{\psi}(\boldsymbol{r}) \label{eq:charge}
\end{equation}
where $\hat{\psi}^{(\dagger)}(\boldsymbol{r})$ is the electron field annihilation (creation) operator, which satisfies the Fermi anti-commutation relation $\{\hat{\psi}(\boldsymbol{r}),\hat{\psi}^{\dagger}(\boldsymbol{r}') \}=\delta (\boldsymbol{r}-\boldsymbol{r}')$.
The vector potential is written as a field mode expansion
\begin{equation}
\hat{\boldsymbol{A}}(\boldsymbol{r})=\sum_{\boldsymbol{k}_j \lambda_j}\sqrt{\frac{2\pi}{\Omega \omega_j}} \bigg( \epsilon^{(\lambda_j)}(\boldsymbol{k}_j)\hat{a}_j e^{i\boldsymbol{k}_j \cdot \boldsymbol{r}} + \epsilon^{(\lambda_j)*}(\boldsymbol{k}_j)\hat{a}^{\dagger}_j e^{-i\boldsymbol{k}_j \cdot \boldsymbol{r}} \bigg)
\label{eq:vectorpotential}
\end{equation}
where $\hat{a}_j^{(\dagger)}$ is the photon field boson annihilation (creation) operator for mode $j$, $\Omega$ the field quantization volume, and $\epsilon^{(\lambda_j)}(\boldsymbol{k}_j)$ the polarization vector.
% The charge density operator is given by $\hat{\sigma}=e|\boldsymbol{r}\rangle\langle\boldsymbol{r}|$.
In the minimal-coupling Hamiltonian, the exact light-matter coupling can be obtained by the substitution $\hat{\boldsymbol{p}} \rightarrow \hat{\boldsymbol{p}}-e\hat{\boldsymbol{A}}$, where $\hat{\boldsymbol{p}}$ is the electronic momentum operator.
The matter property enters through the current $\hat{\boldsymbol{j}}(\boldsymbol{r})$ and charge density $\hat{\sigma}(\boldsymbol{r})$ operators, and the light property through the vector potential $\hat{\boldsymbol{A}}$.
In this formalism, an off-resonant Raman process is described by the transition charge density (TCD) $\sigma_{ij}(\boldsymbol{r})$, which can be calculated as a transition property between the states $i$ and $j$.
A resonant transition is described by the transition current density $\boldsymbol{j}_{ij}(\boldsymbol{r})$.
In the multipolar Hamiltonian, on the other hand, the off-resonant Raman transition between the states $i$ and $j$ is described by the transition polarizability $\alpha_{ij}$, which requires additional computational cost to sum over all the relevant intermediate electronic states $k$. % $k$ $\mu_{ik}\mu_{kj}$
The minimal-coupling Hamiltonian approach is, therefore, more suitable for the calculation of an off-resonant Raman process than the multipolar Hamiltonian.
Simulation of hard X-ray spectroscopy also requires a description of the spatial variation of the field across the molecular sample, and this is also most suited to being calculated in the minimal-coupling formalism.

\begin{figure}[!h]
\includegraphics[width=0.9\textwidth]{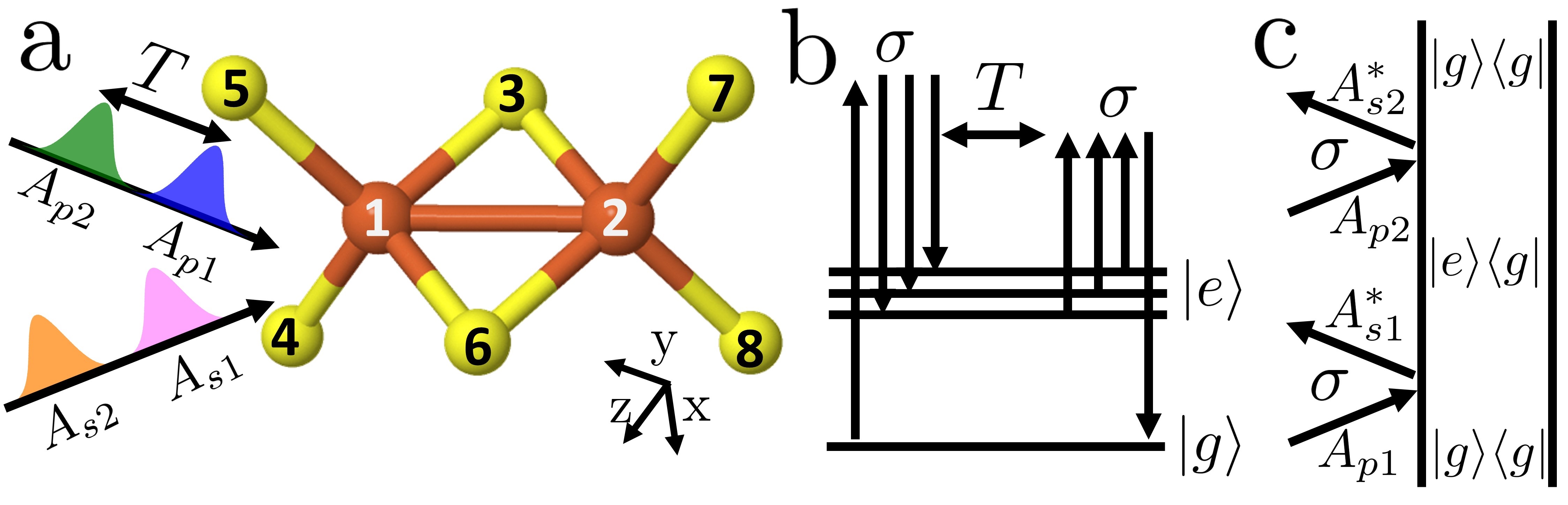}
\caption{(a) Sketch of the pulse configuration relative to the [2Fe-2S] complex. (b) Level diagrams representing the process. $\ket{g}$ and $\ket{e}$ are the ground
and valence excited states. (c) Ladder diagram of the signal.}
\label{scheme}
\end{figure}

In the setup shown in Fig. \ref{scheme}, two X-ray or UV pulses with a controlled delay $T$ are incident on a sample (Fig. \ref{scheme}(a)) and induce two Raman processes (Fig. \ref{scheme}(b)). The signal is measured as a function of $T$ and then Fourier transformed to obtain a spectrum which reveals the valence manifold.  Fig. \ref{scheme}(c) gives the ladder diagram of the signal.

By neglecting the  $\boldsymbol{j}\cdot\boldsymbol{A}$
term, which dominates resonant scattering, the off-resonant heterodyne signal is defined as the time-integrated rate of change of photon number of the field mode occupied by the heterodyne pulse $\boldsymbol{A}_{s2}$ (Fig. \ref{scheme}c)
\begin{equation}
S_{\text{SXRS}}=\int dt \left\langle \frac{d}{dt} \hat{N}_{s2}(t) \right\rangle
\label{eq:SXRS_photon_number}
\end{equation}
where the number operator for the photon mode with heterodyne field $\hat{N}_{s2}=a^{\dagger}_{s2}a_{s2}$ and $a^{(\dagger)}_{s2}$ is a photon annihilation (creation) operator in the $s2$ field mode.
%{\color{red}It seems asymmetrical to define the second-quantized form of the number operator and not that of the vector potential, since both are required to do the next step of computing the commutator.}
Computing the commutator $[\hat{H}_\text{int},\hat{N}_{s2}]$ then gives the signal as %{\color{red}This is just my ignorance, but I assume you are writing $A = A_s + A_p$, then only keeping some terms corresponding to the process in the diagram of the process ...}
%{\color{blue} Daeheum: This was explained by saying that "time-integrated rate of change of photon number of the field mode occupied by the heterodyne pulse A2"}
\begin{equation}
{\normalcolor {\normalcolor }}S_{\text{SXRS}}(\boldsymbol{k}_s)=-\frac{2}{\hbar}\Big(\frac{-e}{2m}\Big)\Im\int d\boldsymbol{r}dt\langle\hat{\sigma}(\boldsymbol{r},t)\rangle\boldsymbol{A}_{s2}^{*}(\boldsymbol{r},t)\cdot\boldsymbol{A}_{p2}(\boldsymbol{r},t)
\label{eq:SXRS_General}
\end{equation}
where $\boldsymbol{k}_s$ is the wavevector of the signal field.
$\boldsymbol{A}_{p2}(\boldsymbol{r},t)$ and $\boldsymbol{A}_{s2}^{*}(\boldsymbol{r},t)$ are the probe field and its scattered field, respectively as shown in Fig. \ref{scheme}(c). Similarly, the pump fields to create a $\kb{g}{e}$ coherence are labelled $\boldsymbol{A}_{p1}(\boldsymbol{r},t)$ and $\boldsymbol{A}_{s1}^{*}(\boldsymbol{r},t)$ as in the Figure.
The off-resonant SXRS is obtained by expanding eq. \ref{eq:SXRS_General} to second
order in $\sigma\boldsymbol{A}^{2}$ and keeping the terms corresponding to the process in Fig. \ref{scheme}(c),
\begin{multline}
S_{\text{SXRS}}(\boldsymbol{k}_s)=-\frac{2}{\hbar}\Big(\frac{-e}{2m}\Big)^{2}\Im(-\frac{i}{\hbar})\int d\boldsymbol{r}_{2}d\boldsymbol{r}_{1}dt_{2}dt_{1}\ \\
\langle \hat{\sigma}(\boldsymbol{r}_{2},t_{2})\hat{\sigma}(\boldsymbol{r}_{1},t_{1})\rangle\boldsymbol{A}_{s2}^{*}(\boldsymbol{r}_{2},t_{2})\cdot\boldsymbol{A}_{p2}(\boldsymbol{r}_{2},t_{2})\boldsymbol{A}_{s1}^{*}(\boldsymbol{r}_{1},t_{1})\cdot\boldsymbol{A}_{p1}(\boldsymbol{r}_{1},t_{1}).
\label{eq:SXRS_correlation_function}
\end{multline}
The incoming fields have a plane wave form $\boldsymbol{A}_{pi}(\boldsymbol{r}_{i},t_{i})
=\mathcal{A}_{pi}\bm{\epsilon}_{i}
e^{i(\boldsymbol{k}_{pi}\cdot r-\omega_{pi}t)}$
where $\mathcal{A}_{pi}$ is the field amplitude, $\bm{\epsilon}_{i}$
its polarization, $\bm{k}_{pi}$ its wavevector and $\omega_{pi}$
its frequency. The signals are computed for the $X$, $Y$, $Z$ incoming pulse directions.
The two-point correlation function of charge densities in eq. \ref{eq:SXRS_correlation_function}, $\langle\sigma(\boldsymbol{r}_{2},t_{2})\sigma(\boldsymbol{r}_{1},t_{1})\rangle$ can be dissected into two contributions: a two-molecule contribution and a one-molecule one.
The first term gives rise to a structure factor as a prefactor that vanishes in the absence of order. The one-molecule contribution does not vanish upon rotational averaging and is the expression used for gas, liquid phase or single molecule scattering.
Note that our simulations will assume an oriented single-molecule (the relative orientation with the fields is shown in Fig \ref{scheme}(a)). This single-molecule orientation in the gas phase may be prepared by an extra aligning pulse prior to an SXRS measurement.
To describe gas or liquid phase signals without molecular ordering, rotational averaging should otherwise be performed. Some structural information will be lost upon rotational averaging over such randomly oriented molecules, but the main spectral features should be similar.

After expanding the matter correlation function in eigenstates, assuming that all incoming pulses have the same polarization we obtain
\begin{equation}
S_{\text{SXRS}}(\Omega)=\Big(\frac{e}{2m}\Big)^{2}\frac{2}{\hbar^{2}}\Im\sum_{e}\Bigg(\frac{\sigma_{ge}(\boldsymbol{q}_{2})\sigma_{eg}(\boldsymbol{q}_{1})}{\Omega-\omega_{eg}-i\Gamma_{eg}}+\frac{\sigma_{ge}(-\boldsymbol{q}_{2})\sigma_{eg}(-\boldsymbol{q}_{1})}{\Omega+\omega_{eg}-i\Gamma_{eg}}\Bigg)
\label{eq:1D_SXRS_sigma}
\end{equation}
where $\Omega$ is the Fourier variable conjugate to the time delay $T$, $\sigma_{ge}(\boldsymbol{q})$ is the Fourier transform
of the transition charge density $\sigma_{ge}(\boldsymbol{r})$, the momentum transfer $\boldsymbol{q}_{i}=\boldsymbol{k}_{si}-\boldsymbol{k}_{pi}$, and the dephasing rate $\Gamma_{eg}=0.014$ eV,.
In the long wavelength limit we have
\begin{multline}
\sigma_{eg}(\boldsymbol{q}_{i})=\int e^{i\boldsymbol{k}_{si}\cdot\boldsymbol{r}}e^{-i\boldsymbol{k}_{pi}\cdot\boldsymbol{r}}\sigma_{eg}(\boldsymbol{r})d\boldsymbol{r}=\int\sigma_{eg}(\boldsymbol{r})(1+i\boldsymbol{k}_{si}\cdot\boldsymbol{r})(1-i\boldsymbol{k}_{pi}\cdot\boldsymbol{r})\\
=\int\sigma_{eg}(\boldsymbol{r})d\boldsymbol{r}+i\int(\boldsymbol{k}_{si}-\boldsymbol{k}_{pi})\boldsymbol{r}\sigma_{eg} d\boldsymbol{r} + \int d\boldsymbol{r}(\boldsymbol{k}_{si}\cdot\boldsymbol{r})(\boldsymbol{k}_{pi}\cdot\boldsymbol{r})\sigma_{eg}(\boldsymbol{r}).\label{eq:}
\end{multline}
The first term vanishes by the definition of the transition charge
density.
The magnitude of the difference between the  in the second term ($\boldsymbol{k}_{si} - \boldsymbol{k}_{pi}$) is small compared to that of the product appearing in the last term ($\boldsymbol{k}_{si} \cdot \boldsymbol{k}_{pi}$) and can thus be neglected. We then obtain for the effective transition charge density.
\begin{equation}
\boldsymbol{\alpha}_{eg}=\frac{e}{2m\hbar c^{2}}\omega_{si}\omega_{pi}\int(\boldsymbol{\epsilon}_{si}\cdot\boldsymbol{r})(\boldsymbol{\epsilon}_{pi}\cdot\boldsymbol{r})\sigma_{eg}(\boldsymbol{r})d\boldsymbol{r}\label{eq:Alpha}
\end{equation}
where $\boldsymbol{\epsilon}_{si}$ and $\boldsymbol{\epsilon}_{pi}$ are the direction of propagation of the $i$th probe and the scattered fields, respectively. The effective TCD $\boldsymbol{\alpha}_{eg}$ no longer depends on $\boldsymbol{r}$ since we have made the long-wavelength approximation. The signal in eq. \ref{eq:1D_SXRS_sigma} is given by
\begin{equation}
S_{\text{SXRS}}(\Omega)=2\Im\sum_{e}\Bigg(\frac{\boldsymbol{\alpha}_{ge}\boldsymbol{\alpha}_{eg}}{\Omega-\omega_{eg}-i\Gamma_{eg}}+\frac{\boldsymbol{\alpha}_{ge}\boldsymbol{\alpha}_{eg}}{\Omega+\omega_{eg}-i\Gamma_{eg}}\Bigg).\label{sigdef}
\end{equation}
In this paper, we take all the pulse propagation axes  $\boldsymbol{\epsilon}_{si}$ and $\boldsymbol{\epsilon}_{pi}$ the same for the SXRS signal, either X, Y, or Z axis shown in Fig. \ref{scheme}.

The absorption spectrum $S_{\text{L}}$ was calculated from the
transition dipole moment $\boldsymbol{\mu}_{eg}$ between states $g$ and $e$,
transition frequency $\omega_{eg}$, and the dephasing rate $\Gamma_{eg}=0.014$ eV,
\begin{equation}
S_{\text{L}}(\omega)=\sum_{e}\frac{|\bm{\mu}_{eg}|^{2}\Gamma_{eg}}{(\omega-\omega_{eg})^{2}+\Gamma_{eg}^{2}}.\label{eq:S_L}
\end{equation}

{\it Computational methods for the electronic excited states of [2Fe-2S] complexes}
The computation of the electronic structure of Fe-S complexes is challenging due to the presence of many nearly degenerate d orbitals. A minimal active space for the ground state, which includes the 3d orbitals of Fe
and 3p orbitals of S (to qualitatively capture double exchange~\cite{anderson1955considerations}) already
contains 16 orbitals, which  approaches the limit of
traditional multi-reference methods.
To more accurately describe the low-lying excited states by allowing for the effects of d-electron orbital relaxation, a second set of d-shell orbitals~\cite{veryazov2011select} was added into the active space, along with two 4s orbitals as in previous work\cite{sharma_low-energy_2014}. In addition, the eight terminal thiolate p orbitals (one $\sigma$ and $\pi$ each\cite{chilkuri2019ligand}) were included as well. This gives rise to complete active space (CAS) models of size CAS(38e,36o) for the ferric-ferric dimer and CAS(39e,36o) for the
ferric-ferrous dimer. These are expanded from the
previously employed active spaces in Ref. \cite{sharma_low-energy_2014} by the
four $\pi$ orbitals from the thiolates, which are included here so as to allow for possible ligand-to-metal charge-transfer (LMCT) in
the excitations.
Note that because empty sulfur orbitals are not
included, our active space model excludes metal-to-ligand charge-transfer (MCLT) from Fe to S. However, the energies of such excited states are very high (ca. >150,000 cm$^{-1}$)\cite{chilkuri2019ligand}
and thus do not contribute to the low-lying spectra investigated in this work.
Thus, the current active space may be considered to provide a qualitative
model of the low-lying excited states, capturing both the d-d
as well as low-lying LMCT transitions, which both appear in the
the low-energy spectra of Fe-S complexes. While further inclusion of dynamic correlation is desirable\cite{presti2019full} and will be a
topic for future study, using the current active spaces in conjunction
with computing many excited states already results in challenging calculations.

\begin{figure}[h!]
\begin{tabular}{cc}\centering
\includegraphics[width=0.5\textwidth]{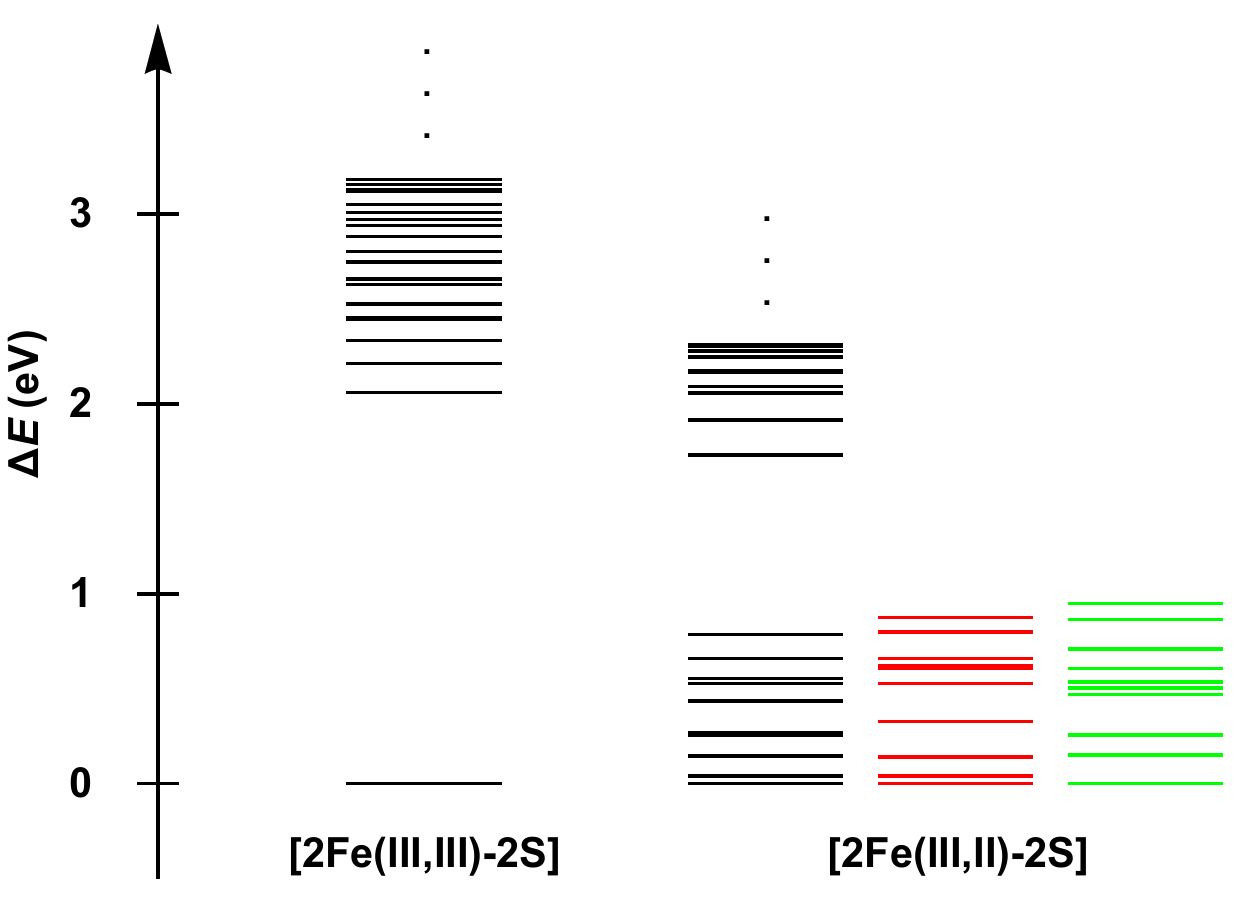} &
\includegraphics[width=0.4\textwidth]{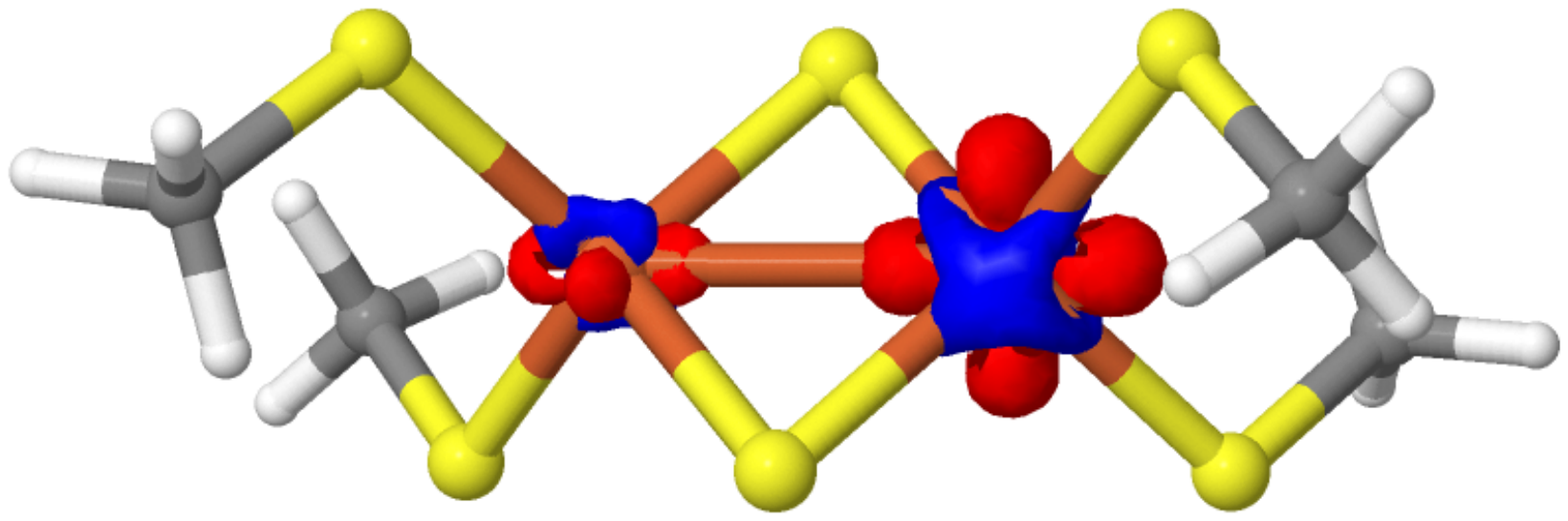} \\
(a) & (b) \\
\end{tabular}
\caption{(a) Relative energies of the 20 low-lying electronic
states of the ferric-ferric dimer with $S=0$ (left) and
the ferric-ferrous (right) dimer with $S=1/2$.
For comparison, the relative energies obtained
in the previous work (Ref. \cite{sharma_low-energy_2014})
were also shown (red: unrelaxed geometry; green: relaxed geometry).
(b) The charge density difference between the ferric-ferrous dimer and the ferric-ferric dimer, which illustrates the distribution of the excess electron.}\label{fig:structure}
\end{figure}

While our previous study\cite{sharma_low-energy_2014}
mostly focused on the lowest 10 states of the mixed-valence dimer, to access a larger part
of the electronic spectrum, we computed 20 electronic states (1 ground state + 19 excited states of the same spin) for each dimer in this work using state-averaged DMRG with
the above enlarged active space and some other
additional improvements described below.
The model complexes for
the [2Fe-2S] clusters were the same as used previously\cite{sharma_low-energy_2014,chilkuri2019ligand},
and were obtained from the synthetic complex of Mayerle et al.\cite{mayerle1973synthetic,mayerle1975synthetic} with the terminal groups simplified to methyl groups in order to reduce computational cost.
The protocol for preparing active space orbitals described in Ref. \cite{li2018electronic} was employed in this work. This is based on split-localization (using Pipek-Mezey (PM) localization\cite{pipek1989fast})
of the unrestricted natural orbitals (UNOs) obtained from high-spin unrestricted Kohn-Sham (UKS) calculations using the BP86\cite{becke1988density,perdew1986density} functional.
Scalar relativistic effects were taken into account by the spin-free exact two-component (sf-X2C) Hamiltonian\cite{liu2010ideas,saue2011relativistic,peng2012exact,li2012spin}, and the cc-pVTZ-DK basis\cite{balabanov2005systematically} was employed
for all atoms. All of these calculations were performed with the PySCF package\cite{sun2017pyscf}. The resulting active orbitals are visualized in Sec. 1 in the supporting information (SI).
The subsequent state-averaged DMRG calculations were performed with the
BLOCK code\cite{sharma_spin-adapted_2012}. Most of the
results (vertical excitation energies, charge and spin densities, and transition density matrices) presented in this paper were obtained from data corresponding to a (spin-adapted) bond dimension $D=2000$.
The vertical excitation energies for the lowest
19 excited states $\omega_{eg}$ (in eV) for the Fe(III)-Fe(III)
dimer with $S=0$ and the Fe(III)-Fe(II) dimer with $S=1/2$ are summarized in Table \ref{tab:Excitation_Character_Fe(III)-Fe(III)-wide} and Fig.
\ref{fig:structure}(a). DMRG calculations averaging over  many states are computationally very expensive. To estimate the errors in our calculations, we compared the results at $D=2000$ with a calculation with larger bond dimension $D=3000$ (only
performing 2 sweeps due to the computational cost).
Because of the variational nature of DMRG, a
lower energy (for a given state) means better convergence than a higher energy.
 For the first 9 excited states, on average, the change in excitation energies was -0.05 eV
and -0.02 eV for the [2Fe(III,III)-2S] and [2Fe(III,II)-2S] complexes,
respectively. For the next 10 higher energy excited states, the average changes were larger (-0.28 eV and -0.23 eV, respectively)
although the qualitative features of the states remained unchanged.
As shown in Fig. \ref{fig:structure}(a),
our results for the relative energies of the lowest 10 states of
the mixed-valence dimer is in agreement with our previous results\cite{sharma_low-energy_2014}.

\begin{table}
\caption{Vertical excitation energies for the lowest
19 excited states $\omega_{eg}$ (in eV) for Fe(III)-Fe(III) ($S=0$)
and Fe(III)-Fe(II) ($S=1/2$) dimers from state-averaged DMRG.
The values in  parentheses correspond to $\mathrm{tr}[\gamma^\dagger \gamma]$, where $\gamma$ is the transition density matrix.
The deviation from one can be regarded as a signature that the excitation involves multiple (instead of single) excitation character.}\scriptsize
\begin{tabular}{c|cccccccccc}
\hline\hline
\multirow{2}{*}{\diagbox{complex}{state}} & 1 & 2 & 3 & 4 & 5 & 6 & 7 & 8 & 9 & 10 \\ & 11 & 12 & 13 & 14 & 15 & 16 & 17 & 18 & 19 \\
\hline
Fe(III)-Fe(III)
& 2.06 & 2.21 & 2.33 & 2.44 & 2.45 & 2.53 & 2.63 & 2.66 & 2.75 & 2.80 \\
& (0.54) & (0.63) & (0.62) & (0.64) & (0.66) & (0.68) & (0.65) & (0.66) & (0.62) & (0.60) \\
& 2.88 & 2.94 & 2.97 & 3.01 & 3.05 & 3.12 & 3.13 & 3.16 & 3.18\\
& (0.62) & (0.62) & (0.67) & (0.58) & (0.62) & (0.65) & (0.66) & (0.54) & (0.62) \\
Fe(III)-Fe(II)
& 0.04 & 0.15 & 0.26 & 0.27 & 0.44 & 0.53 & 0.55 & 0.66 & 0.78 & 1.73 \\
& (0.45) & (0.47) & (0.46) & (0.53) & (0.47) & (0.61) & (0.64) & (0.26) & (0.34) & (0.24) \\
& 1.91 & 2.06 & 2.09 & 2.17 & 2.18 & 2.25 & 2.28 & 2.30 & 2.31 \\
& (0.07) & (0.23) & (0.38) & (0.20) & (0.06) & (0.11) & (0.18) & (0.11) & (0.38) \\
\hline\hline
\end{tabular}
\label{tab:Excitation_Character_Fe(III)-Fe(III)-wide}
\end{table}

{\it Nature of the low-lying excited states}
Before presenting the simulated spectra, we first analyze the nature of the low-lying excited states, which gives some basic insight into the electronic structure of the low-lying states. Figure 2(a) reveals that
for the ferric-ferric dimer, there is a single dense band
around 2-3 eV formed by the first 19 excited states,
whereas for the ferric-ferrous dimer, the computed excited states
split into two bands. The first 10 states (including the ground
state) form a single band within 0.78 eV, while the next
band formed by the rest of the states starts from 1.73 eV.

To analyze the excited states, in Table \ref{tab:Excitation_Character_Fe(III)-Fe(III)-wide},
the values of $\mathrm{tr}[\gamma^\dagger \gamma]$, where
$\gamma$ is the one-particle transition density matrix
defined as $(\gamma_{ge})_{pq}=\langle\Psi_g|a_p^\dagger a_q|\Psi_e\rangle$,
are listed for each excited state. Significant deviation of this value from one is a sign that multiple (instead of single) excitations are
involved in $|\Psi_e\rangle$\cite{matsika2014we,hu2015excited,ren2017role}. (In the single-reference case, the above statement is exactly true, because if $\Psi_g$ is described by a Slater determinant and $|\Psi_e\rangle$ is described at the level of CIS (configuration interaction singles), the corresponding
value is precisely one, i.e., $\mathrm{tr}[\gamma^\dagger \gamma]=1$).
As shown in Table \ref{tab:Excitation_Character_Fe(III)-Fe(III)-wide},
a common feature for the low-lying excited states in both Fe-S clusters is that they all contain substantial multiple excitation character.
In particular, the excited states in the second band of the ferric-ferrous dimer can even be considered to be dominated by multiple excitations, which is also the case for the 8th and 9th excited states in the first band.

The origin of the excited states can be analyzed by visualizing the charge density differences between the excited states and the ground state
shown in Figs. S2 and S3. For the ferric-ferric dimer, Fig. S2 shows the deletion of electron density on the bridge sulfur orbitals and the rearrangement of electron density on the two Fe ions. This indicates that the dense band of excited states for the ferric-ferric dimer can be attributed to a strong mixture of d-d excitations and LMCT from
the bridging sulfur to the Fe ions. In contrast, for the reduced dimer,
Fig. S3 reveals that the excited states are mainly composed of
d-d excitations. Only very few of them involve a small amount
of LMCT (e.g., see the 10th, 11th, 14th, 15th, and 19th states in Fig. S3).

The exact nature of the d-d excitations is very hard to discern
due to the heavy mixture of different types of single and multiple d-d excitations in the iron-sulfur dimers. There can be (A) local d-d excitations within one center, (B) simultaneous local d-d excitations on both centers,
(C) charge-transfer d-d excitations between two centers,
and multiple excitations with mixed character. Both  types (A) and (C) of d-d excitations can either be single or multiple transitions, whereas
the other types are multiple excitations by definition. From the charge density difference alone,
it is difficult to trace the structure of the d-d excitations in the computed excited states to these different classes.
Fortunately, for the purpose of understanding the spectroscopies shown below, it suffices to focus on the most relevant quantity, the TCD shown in Fig. S4, for which only single excitations (types (A) and (C)) are relevant. To obtain more compact information, we decompose the transition density matrix
$\gamma$ using a singular value decomposition, $\gamma=\mathbf{U}\Lambda \mathbf{V}^\dagger$ in a way similar
to the definition of natural transition orbitals (NTO) in the case of CIS\cite{martin2003natural}. The resulting pairs of orbitals
are referred to as  binatural orbitals in the multi-reference
context\cite{malmqvist2012binatural}. For simplicity, the two sets of orbitals defined by $\mathbf{U}$ and $\mathbf{V}$ will be denoted hole NTO (HNTO) and electron NTO (ENTO), respectively.
The contribution of each pair of HNTO and ENTO to
the transition density matrix is given by the singular value
$\lambda_{k}$. However, as mentioned above, in the multireference
case, one usually finds $\mathrm{tr}[\gamma^\dagger \gamma]=\mathrm{tr}(\Lambda^\dagger\Lambda)=\sum_{k}\lambda_k^2< 1$.
Thus, when discussing the contributions of each pair of HNTO and ENTO to the total TCD, we will use the normalized percentage $\lambda_{k}^2/\sum_l\lambda_l^2$ (see Figs.
S5 and S6 in SI). By analyzing the pairs of NTOs, we can interpret the character of the electronic transition contributing to the TCDs in a compact way.

Figure S5 shows the HNTOs and ENTOs for the ferric-ferric dimer.
We find that the TCDs (see Figure S4(a)) for all excited states are mostly contributed by LMCT and d-d excitations with both local and charge-transfer character.
This is consistent with the findings from analyzing charge density differences
and the observation that the low-lying excited states
in this complex contain a relatively larger amount of single excitation
character than multiple excitation character (see Table \ref{tab:Excitation_Character_Fe(III)-Fe(III)-wide}).
For the ferric-ferrous case, Fig. S6 shows that
the TCDs (see Figure S4(b)) of the first band of excited states are purely due to local d-d excitations, viz., mostly
d(Fe2)$\rightarrow$d(Fe2) transitions combined with
a small amount of d(Fe1)$\rightarrow$d(Fe1) transitions.
This is in agreement with the fact that upon reduction of the ferric-ferric cluster, the Fe2 ion becomes
more reduced than the Fe1 ion, as revealed by the charge density
difference shown in Fig. \ref{fig:structure}(b).
The lowest energy absorption from the ground to the first excited state is contributed by the local excitation between
two split $e$ band d-orbitals of the ferrous iron (Fe2)
due to the distorted tetrahedral environment. The small energy splitting of 0.04 eV (about 300 cm$^{-1}$) is in line with the observed energy
splitting (400 cm$^{-1}$) for the ferrous iron in spinach and parsley ferredoxins
estimated from fitting to M{\"o}ssbauer spectra\cite{dunham1971two,dunham1971structure}.
Further, according to Fig. S6, the dominant contributions to the TCDs of the second band of excited states for the ferric-ferrous dimer come from charge-transfer d-d excitations between the ferrous and ferric Fe ions. This is also quite different in nature from the low-lying excited states of the homovalent dimer.

\begin{figure}[h!]
\includegraphics[width=1\textwidth]{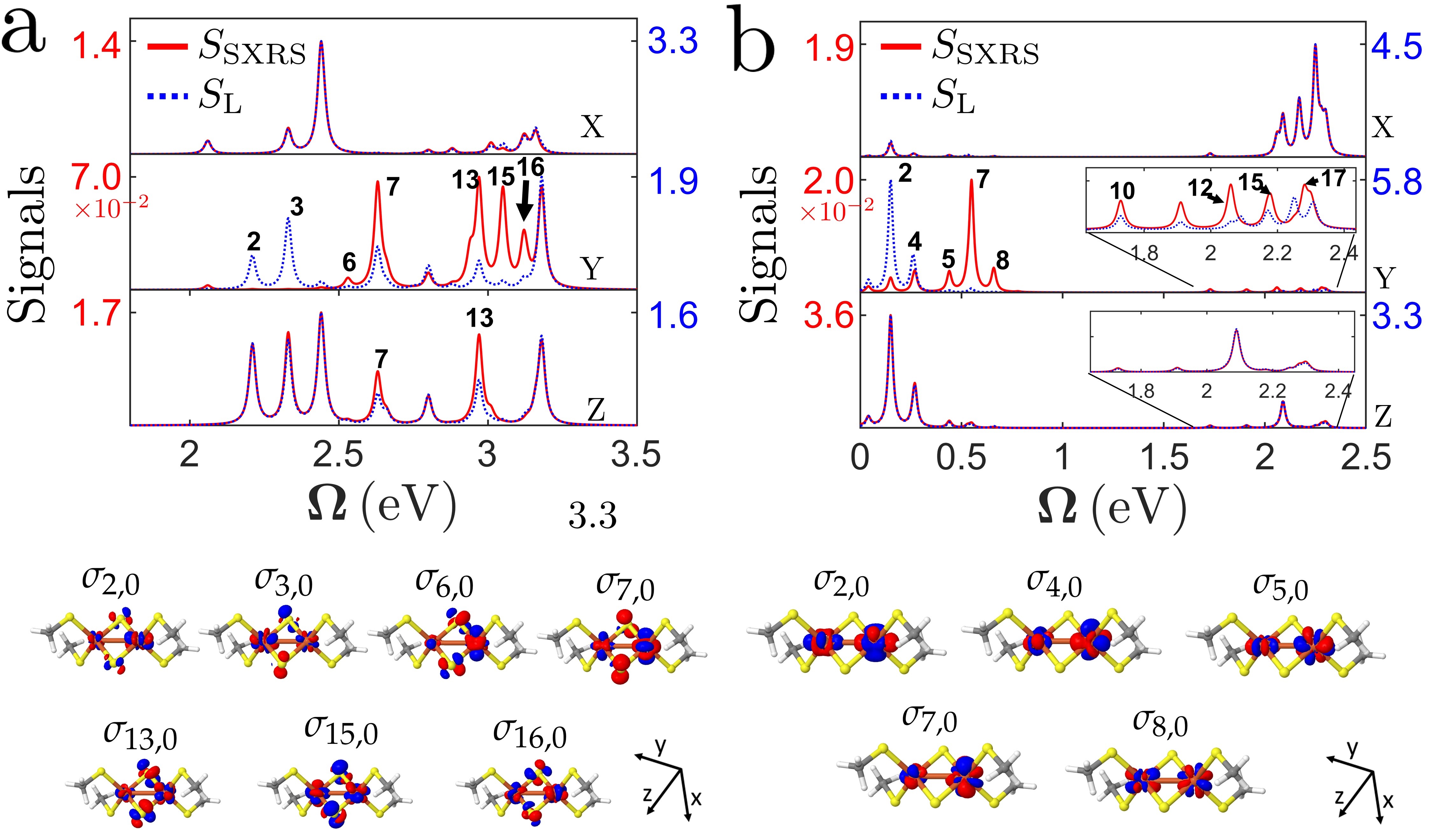} \caption{(a) SXRS signals $S_{\text{SXRS}}$ (Eq. \eqref{sigdef}, solid line)
and absorption signals $S_{\text{L}}$ (Eq. \eqref{eq:S_L}, dashed
line) of (a) Fe(III)-Fe(III) and (b) Fe(III)-Fe(II) dimers. (Top)
Calculated signals from X, Y, and Z polarized light. (See bottom of
Figure b for the axes).  (Bottom) Selected TCDs are shown. $\Gamma_{eg}=0.014$\,eV
for all states. Note that the absorption and SXRS signals are normalized. Consequently, the signal strength of each spectroscopy
in different directions can be compared, but SXRS and absorption strengths cannot be directly compared.}\label{fig:S_SXRS_TOTAL}
\end{figure}

{\it Simulated spectra for iron-sulfur dimers}
We simulated the SXRS signals $S_{\text{SXRS}}$ and absorption signals $S_{\text{L}}$  in X, Y, and Z polarized light for an oriented (e.g. in solid phase) iron-sulfur cluster, see Fig. \ref{fig:S_SXRS_TOTAL}.
Figure \ref{fig:S_SXRS_TOTAL}(a) depicts the normalized signals of the [2Fe(III,III)-2S] dimer, covering a spectral range from 2.06 to 3.18 eV (corresponding wavelength of 601 nm to 389 nm).
The two signals show significant
differences for the Y polarization, but are almost identical for X and Z polarizations. This feature indicates that both absorption and SXRS signals are dependent on the incoming pulse directions.
The absorption almost exclusively spans the 2.2-2.4 eV regime whereas the SXRS spans the 2.9-3.1 eV regime. We find significant signal enhancements in the SXRS for the
6th, 7th, 13th, 15th, and 16th excited states compared to the
absorption. The relative signal enhancements in SXRS
indicates that this technique may allow a better observation of
dark states in the absorption. The separation of the absorption-active (2.2-2.4 eV) and the Raman-active (2.9-3.1 eV) states should allow for the selective excitation of excited states by properly tuning the excitation bandwidth. The ensuing electronic dynamics may then report on the initial electronic superposition and be sensitive to the biochemical environment.

Figure \ref{fig:S_SXRS_TOTAL}(b) depicts the normalized $S_{\text{SXRS}}$ and $S_{\text{L}}$ signals of the Fe(III)-Fe(II) dimer, spanning the 0.04 to 2.31 eV spectral range (corresponding wavelength of 31 $\mu m$ to 536 nm). Note that the low-energy excitation signals are stronger than those in the visible range in the Y and Z polarizations. While the $S_{\text{SXRS}}$ and $S_{\text{L}}$ signals for the second band of excited states
are quite similar, there are significant differences for
the first band of excited states in the Y direction.
The 2nd and 4th states are almost exclusively observed in the absorption, while the 5th, 7th, and 8-th excited states by the Raman excitation in Y polarization. Besides, it is noted that the low energy d-d transitions enable the absorption at longer wavelength light from near infrared (0.78 eV, 1550 nm) down to the mid infrared (microwave) regime (0.04 eV, 31 $\mu m$). In reality, however, the low-energy part of the electronic spectrum and the vibrational spectrum may overlap. This highlights the need for a quantum calculation of a vibronic spectrum, with a proper treatment of molecular vibrations, in the iron-sulfur complexes.

{\it Conclusions} In this work, we employed the high-level {\it ab initio} DMRG algorithm to calculate the low-lying electronic states of [2Fe-2S] dimers in different oxidation states. Consistent with earlier proposals, the reduced dimer exhibits very low-energy electronic excitations
below visible wavelengths. Using the excited states and the transition charge densities, we simulated the off-resonant SXRS and absorption
signals of the dimers. We find significant differences in signal intensity between the absorption-active and the Raman-active states of the iron-sulfur
dimers along one of the axes of incidence, providing a novel means to access previously dark states.
This difference in signal intensity also allows for the selective excitation of excited states by a proper tuning of the excitation bandwidth, thus
probing different types of dynamics following the preparation of an initial electronic superposition. This will be a topic of future work.

%%%%%%%%%%%%%%%%%%%%%%%%%%%%%%%%%%%%%%%%%%%%%%%%%%%%%%%%%%%%%%%%%%%%%
%% The "Acknowledgement" section can be given in all manuscript
%% classes.  This should be given within the "acknowledgement"
%% environment, which will make the correct section or running title.
%%%%%%%%%%%%%%%%%%%%%%%%%%%%%%%%%%%%%%%%%%%%%%%%%%%%%%%%%%%%%%%%%%%%%
\begin{acknowledgement}
  S. M. gratefully acknowledges the the support of the National Science Foundation (grant CHE1663822) and the Chemical Sciences, Geosciences, and Biosciences division, Office of Basic Energy Sciences, Office of Science, U.S. Department of Energy through award No. DE-FG02-04ER15571. D.C was supported by the DOE award. Work performed at Caltech was supported by the US National Science Foundation (CHE1665333). The BLOCK and PySCF programs were developed with
  the support of the US National Science Foundation (CHE1657286). G.K.C. is a Simons Investigator in Physics.
Z.L. acknowledges the Beijing Normal University Startup Package.
\end{acknowledgement}

%%%%%%%%%%%%%%%%%%%%%%%%%%%%%%%%%%%%%%%%%%%%%%%%%%%%%%%%%%%%%%%%%%%%%
%% The same is true for Supporting Information, which should use the
%% suppinfo environment.
%%%%%%%%%%%%%%%%%%%%%%%%%%%%%%%%%%%%%%%%%%%%%%%%%%%%%%%%%%%%%%%%%%%%%
\begin{suppinfo}
Additional computational details and results including 1) active orbitals;
2) charge density differences between excited states and the ground state;
3) transition charge densities; 4) natural transition orbitals for
each excited state of the two iron-sulfur dimers.
\end{suppinfo}

%%%%%%%%%%%%%%%%%%%%%%%%%%%%%%%%%%%%%%%%%%%%%%%%%%%%%%%%%%%%%%%%%%%%%
%% The appropriate \bibliography command should be placed here.
%% Notice that the class file automatically sets \bibliographystyle
%% and also names the section correctly.
%%%%%%%%%%%%%%%%%%%%%%%%%%%%%%%%%%%%%%%%%%%%%%%%%%%%%%%%%%%%%%%%%%%%%
%\bibliography{references}

\providecommand*\mcitethebibliography{\thebibliography}
\csname @ifundefined\endcsname{endmcitethebibliography}
  {\let\endmcitethebibliography\endthebibliography}{}

%%%%%%%%%%%%%%%%%%%%%%%%%%%%%%%%%%%%%%%%%%%%%%%%%%%%%%%%%%%%%%%%%%%%%
%% The "tocentry" environment can be used to create an entry for the
%% graphical table of contents.
%%%%%%%%%%%%%%%%%%%%%%%%%%%%%%%%%%%%%%%%%%%%%%%%%%%%%%%%%%%%%%%%%%%%%

\begin{tocentry}
\includegraphics[width=0.9\textwidth]{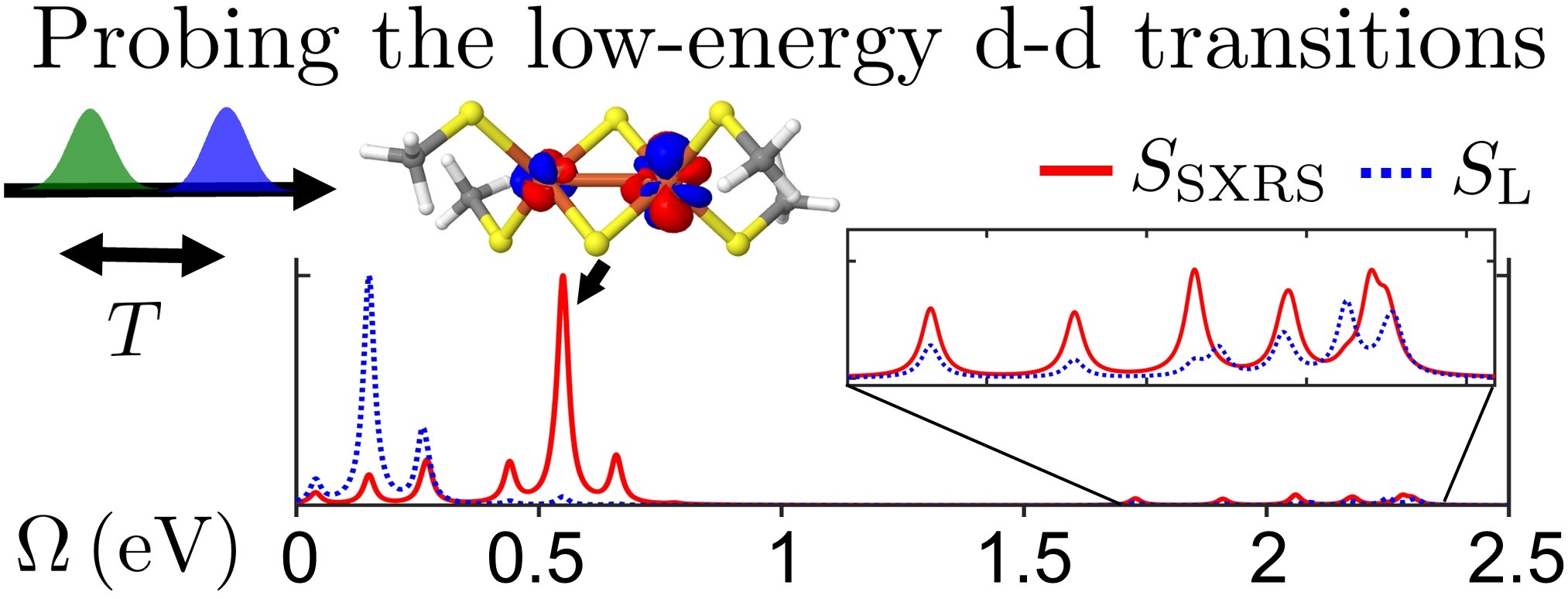}\\
\end{tocentry}

%\begin{figure}[!h]
%\caption{(a) Sketch of the pulse configuration towards the [2Fe-2S] complex. (b) Level diagrams representing the process. $\ket{g}$ and $\ket{b}$ are the ground
%and valence excited states. (c) Ladder diagram of the signal.}
%\label{TOC}
%\end{figure}

% journals require a graphical entry for the Table of Contents.
%This should be laid out ``print ready'' so that the sizing of the
%text is correct.

%Inside the \texttt{tocentry} environment, the font used is Helvetica
%8\,pt, as required by \emph{Journal of the American Chemical
%Society}.

%The surrounding frame is 9\,cm by 3.5\,cm, which is the maximum
%permitted for  \emph{Journal of the American Chemical Society}
%graphical table of content entries. The box will not resize if the
%content is too big: instead it will overflow the edge of the box.

%This box and the associated title will always be printed on a
%separate page at the end of the document.

\end{document}